\documentclass[a4paper]{article}

\usepackage[english]{babel}
\usepackage[utf8x]{inputenc}
\usepackage[T1]{fontenc}

\usepackage[a4paper,top=3cm,bottom=2cm,left=3cm,right=3cm,marginparwidth=1.75cm]{geometry}

\usepackage{amsmath}
\usepackage{algpseudocode}
\usepackage{graphicx}
\usepackage{epsfig}
\usepackage{float}
\usepackage{makecell}
\usepackage{subcaption}
\usepackage[colorinlistoftodos]{todonotes}
\usepackage[colorlinks=true, allcolors=blue]{hyperref}

\title{CoHSI IV:  Unifying Horizontal and Vertical Gene Transfer - is Mechanism Irrelevant ?}
\author{Les Hatton\footnote{Emeritus Professor, Kingston University, KT1 2EE, U.K., lesh@oakcomp.co.uk}, Gregory Warr\footnote{Emeritus Professor, Medical University of South Carolina, 171 Ashley Ave, Charleston, SC 29425, USA, gwawarr@gmail.com}}

\begin{document}

\maketitle

\begin{abstract}
In previous papers we have described with strong experimental support, the organising role that CoHSI (Conservation of Hartley-Shannon Information) plays in determining important global properties of all known proteins, from defining the length distribution, to the natural emergence of very long proteins and their relationship to evolutionary time.  Here we consider the insight that CoHSI might bring to a different problem, the distribution of \textit{identical} proteins across species.  Horizontal and Vertical Gene Transfer (HGT/VGT) both lead to the replication of protein sequences across species through a diversity of mechanisms some of which remain unknown.  In contrast, CoHSI predicts from fundamental theory that such systems will demonstrate power law behavior independently of any mechanisms, and using the Uniprot database we show that the global pattern of protein re-use is emphatically linear on a log-log plot (adj. $R^{2} = 0.99, p < 2.2 \times 10^{-16}$ over 4 decades); i.e. it is extremely close to the predicted power law.  Specifically we show that over 6.9 million proteins in TrEMBL 18-02 are re-used, i.e. their sequence appears identically in between 2 and 9,812 species, with re-used proteins varying in length from 7 to as long as 14,596 amino acids. Using (DL+V) to denote the three domains of life plus viruses, 21,676 proteins are shared between two (DL+V); 22 between three (DL+V) and 5 are shared in all four (DL+V).  Although the majority of protein re-use occurs between bacterial species those proteins most frequently re-used occur disproportionately in viruses, which play a fundamental role in this distribution.

These results suggest that diverse mechanisms of gene transfer (including traditional inheritance) are \textit{irrelevant} in determining the global distribution of protein re-use. 
\end{abstract}

\section*{Statement of reproducibility}
This paper adheres to the transparency and reproducibility principles espoused by \cite{Popper1959,Ziolkowski1982,Claerbout1992,HatRob94,Donoho2009,Ince2012} and includes references to all methods and source code necessary to reproduce the results presented.  These are referred to here as the \textit{reproducibility deliverables} and will be available initially at https://leshatton.org.  Each reproducibility deliverable allows all results, tables and diagrams to be reproduced individually for that paper, as well as performing verification checks on machine environment, availability of essential open source packages, quality of arithmetic and regression testing of the outputs \cite{HattonWarr2016}.  Note that these packages are designed to run on Linux machines for no other reason than to guarantee the absence of any closed source and therefore potentially opaque contributions to these results.
\section*{Introduction}
In its simplest form, Horizontal Gene Transfer (HGT) also known as Lateral Gene Transfer (LGT/LTG) involves the movement of genetic material (inferred in this work from the encoded proteins) between organisms, in contrast to transmission through inheritance, or Vertical Gene Transfer (VGT).  In other words, in HGT the same protein(s) are found scattered across the classical phylogenetic tree, confounding simple concepts of evolution by descent from a common ancestor. Indeed, in January 2009, the science magazine \textit{New Scientist} on the bicentennial of Darwin's birth and the 150th anniversary of the publication of the \textit{Origin of Species,} went so far as to proclaim on its front cover that "Darwin was wrong", and \cite{Boto2010} noted that a new evolutionary paradigm was needed. The justification for this provocative assertion is that the lateral transfer of genes has been recognized as an important factor in evolution. Quoting from  \cite{Woese2002}: "This is an evolutionary force to be reckoned with, comparable in power and consequence to classical vertical evolutionary mechanisms" and "HGT (and a genetic \textit{lingua franca}) is a necessary condition for the evolution of cell designs". The lateral transfer of cells (or significant portions of their genomes) is considered to have played a role in the origins of the eukaryotic cell, and eukaryotic organelles including the nucleus, mitochondria and photosynthetic plastids \cite{Margulis1975,Koonin2015,Martin2015}. HGT is not restricted to distant evolutionary events - it plays a fundamental role in the ecology, pathogenesis and spread of antibiotic resistance in bacteria \cite{Gyles2014,Vaux2017,Ochman2000} and there are documented examples in eukaryotes \cite{DunningHotopp2011,Hotopp2007,Hotopp2017,Bartolome2009,Gladyshev2008} . However, the argument for widespread HGT in eukaryotes continues to be  controversial \cite{Danchin2016}. 

We note in passing that the process of lateral transfer of information is not restricted to living organisms; it is well-understood in databases, and computer scientists call it a degeneration of a \textit{hierarchical} data structure into a \textit{plex} data structure. 

Mechanisms for HGT in prokaryotes have been described in addition to the well-known phenomena of transformation (the common method of plasmid transfer), transduction (whereby genes are carried by phage) and conjugation (where DNA is directly exchanged between cells) \cite{McDaniel2010,Wagner2017}. Although no general mechanisms have been clearly described for HGT from prokaryotes to eukaryotes or for HGT between eukaryotes (with the exception of retroviral infection) many examples of such HGT are documented (e.g. noted above), and in some cases there is plausible evidence for the involvement of arthropod vectors (\cite{Danchin2016,Walsh2013,Gao2018})

HGT is clearly not a single phenomenon and it lacks a global mechanism; in other words it lacks any obvious unifying principles. Thus it might be useful to take an entirely different approach, one which does not rely on local mechanisms of any kind.  We shall show here that rather than HGT proving "Darwin was wrong", both HGT and VGT are simply subsets of an umbrella concept, \textit{protein re-use} in which \textit{no mechanisms need to be invoked}. Indeed the distribution of protein re-use owes nothing to evolution, it is simply an emergent property of a discrete system constrained by a conservation principle (Conservation of Hartley-Shannon Information) derived from information theory. This takes nothing away from evolution; but protein re-use (which reflects both HGT and VGT) is responsive to a conservation principle acting at a global and therefore deeper level than natural selection.

In previous papers, we have applied global conservation principles as constraints in a Statistical Mechanical model to explain the length distribution of proteins measured in amino acids \cite{HattonWarr2015,HattonWarr2017,HattonWarr2018a}; to explain the remarkable constancy of measurements of average protein length \cite{HattonWarr2018b}; and to explain the inevitable appearance of unusually long proteins \cite{HattonWarr2018c}.  In each case, we have been able to confirm the predictions of this model to high levels of statistical significance.  The model blends the use of the Conservation of Hartley-Shannon Information inside a methodology based on Statistical Mechanics and we will refer to it here as the CoHSI model.

The value of this model in describing discrete systems of very different provenance at all scales is that Hartley-Shannon Information is \textit{token-agnostic.}  When applied to the known collection of all proteins, each protein (a component of the system) can be considered as a string of amino acids (tokens) drawn from a unique alphabet of 22 directly decoded from DNA, supplemented with those which are post-translationally modified (PTM). Token-agnostic means that Hartley-Shannon Information does not care what the tokens actually mean.  It is sensitive only to \textit{changes} of token - and in the case of amino acids, their individual physico-chemical properties are irrelevant.

In addition to this token-agnostic information measure, the methodology of Statistical Mechanics is itself \textit{mechanism-agnostic.}  When applied, it leads to the concept of the overwhelmingly most likely distribution to be observed in a system of a certain total size in tokens T, divided amongst M components (which we defined as proteins), and a certain total Hartley-Shannon Information content I.  No mechanism other than the assumption that all microstates are equiprobable is required and for this reason, we say that Statistical Mechanics is mechanism free.   When one constraint is total energy $E$ rather than total Information $I$, it leads directly to the famous Maxwell-Boltzmann distribution \cite{GlazerWark2001}.

The full details of the blending of Statistical Mechanics with total Hartley-Shannon information content are available in \cite{HattonWarr2017} and the theory leads to the definition of two kinds of discrete systems, the \textit{heterogeneous} case, (where each component contains different kinds of token), and the \textit{homogeneous} case, (where each component contains just one kind of token and no two components contain the same kind of token). Using the token- and mechanism-agnostic methodology of CoHSI in this study we will consider proteins as tokens in the homogeneous discrete system of all known proteins (the TrEMBL database), and predict the expected frequency of occurrence of protein copies across species and domains of life, \textit{without regard to any mechanism by which those copies might have been created.}  We will then compare this prediction with the actual occurrence rates of protein re-use in version 18-02 of the TrEMBL protein database (https://uniprot.org/).

\textit{The fundamental question we address in this study is whether or not protein re-use considered across all domains of life plus the viruses, is constrained by the conservation principle embodied by CoHSI, rendering discussions of mechanisms irrelevant.}

\section*{Methods} 
To re-iterate \textbf{all} software used and methods of deployment are available in the reproducibility package described earlier and accompanying this paper so all results can be reproduced.

\subsection*{Statement of the goal}
We should emphasize first that we qualify as examples of re-use only proteins of \textit{identical sequence}, thus avoiding the potentially complicating issues of mutation,  allelic polymorphism and orthology or paralogy. The statement of the problem is therefore deceptively simple.

\textit{How often and where do proteins with exactly the same sequence of amino acids appear in the full TrEMBL database?}  This presents some significant data processing challenges.  As genetic sequencing and ancillary methods become faster, more affordable and consequently more ubiquitous, standard protein repositories such as TrEMBL \cite{Trembl2018} are currently growing exponentially, even after a significant culling in 2015.  TrEMBL release 18-02, as used here, is already almost 71GB. in its compressed form and has grown by over 40\% in 12 months.  Of course, the databases can be queried online but even relatively simple BLAST queries can take a long time to complete.

In our case, we need to answer specific \textit{global} queries related to protein sequence matching across the entire database, orders of magnitude more quickly.  In particular, we are interested in how often a protein appears \textit{identically} in different species regardless of species relationship or mechanism of gene transfer.  \textit{By identical, we mean precisely the same sequence of amino acids, but without including any possible complications due to PTM (Post-Translational Modification)} as such annotations are currently lagging well behind basic sequencing \cite{Selene2013}. (We note in passing that we would predict this to make no difference, as our methodology is token-agnostic, but this will have to wait.)

To be precise, we wished to find every multiple occurrence of a complete protein sequence of between 2 and nearly 37,000 amino acids in a database comprising (before initial processing) of some 106,206,144 proteins in version 18-02 of TrEMBL, which itself may contain duplicates as it is automatically annotated and not yet reviewed.

As we are studying protein sequences specifically, we term multiple occurrences of an identical protein sequence across species as \textbf{local protein re-use} when it is observed within the same domain of life or entirely within viruses, and \textbf{extended protein re-use} if the identical protein sequence occurs in more than one of the DL+V.  (We use the contraction DL+V rather than the more awkward "three domains of life plus viruses'' whilst the status of viruses as a domain of life remains controversial.)  We note that recent work based on identifying protein folds in some proteins has been used to argue that viruses exchange genes across domains of life \cite{Malik2017}.  Our approach is mechanism-agnostic however so we will simply search for the precise amino acid sequence for each protein across the entire extant TrEMBL database comprising the DL+V, which at the time of writing is version 18-02, and compare them.  We have also carried out a sensitivity analysis to see to what extent single amino acid polymorphisms in the form of changes, omissions or additions, or transpositions between adjacent amino acids in the protein sequence might affect our results, another significant data processing challenge.

\subsection*{Theoretical background}
Some kind of underpinning theoretical model is essential when approaching datasets of this size looking for patterns, otherwise it is all too easy to be overwhelmed by the gigantic and growing complexity.  As we described above, we are using a methodology which accurately predicts amongst other \textit{global} properties, the length distribution of all discrete systems, including proteins and other discrete systems such as software, at all levels of aggregation \cite{HattonWarr2015,HattonWarr2017}.  Although it is described in detail in \cite{HattonWarr2017}, it will be useful to spend a few paragraphs describing exactly how this methodology can be used to approach problems such as determining the pattern of protein re-use without any consideration of specific mechanisms.

\subsubsection*{The CoHSI models}
\begin{figure}[h!]
\centering
\includegraphics[width=8cm,height=6cm]{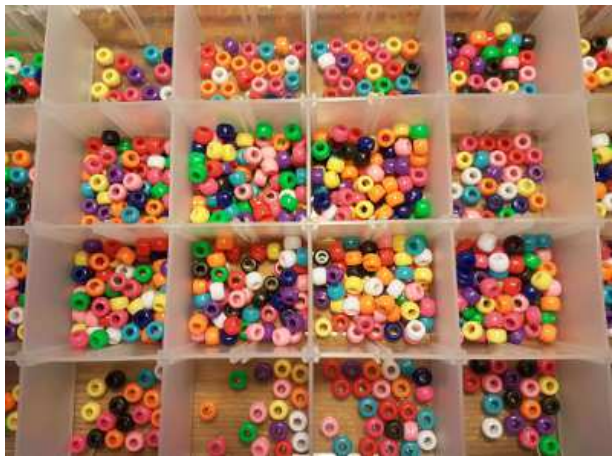}
\caption{Illustrating the CoHSI heterogeneous model.  Each box corresponds to a protein and each of the colored beads is an amino acid.  Different colors indicate different amino acids.  Of course in a protein, they are arranged sequentially and so their order is distinguishable.}
\label{fig:heterogeneous}
\end{figure}

In \cite{HattonWarr2017}, we described two kinds of model for discrete systems.  The first we call the CoHSI (Conservation of Hartley-Shannon Information) \textbf{heterogeneous} model.  Fig. \ref{fig:heterogeneous} illustrates.  In this model, a system is considered to be a set of M components (the boxes) numbered i = 1,..,M. The $i^{th}$ box contains $t_{i}$ beads drawn from a unique alphabet of $a_{i}$ beads, which are here represented as different colors.  There are $T$ beads in total.  We also note that the order of the beads in each box is distinguishable by its position, (imagine them being on a string for example).  We will synonymously use the word \textit{token} for bead in what follows.

This model is appropriate for example for proteins, where each box is a protein and the beads are amino acids, with the colors representing different amino acids.  It is also appropriate for software (where the components are computer functions and the beads are programming language tokens); and also for texts (where each word is a box and the beads are letters).

The important contribution of CoHSI here is that by considering the Hartley-Shannon Information content of each box $I_{i}$ and the system as a whole $I$, it can be shown that the length distribution $t_{i}$ is the solution of the same differential equation \cite{HattonWarr2017} irrespective of the nature of the boxes or beads.

\begin{figure}[h!]
\centering
\includegraphics[width=8cm,height=6cm]{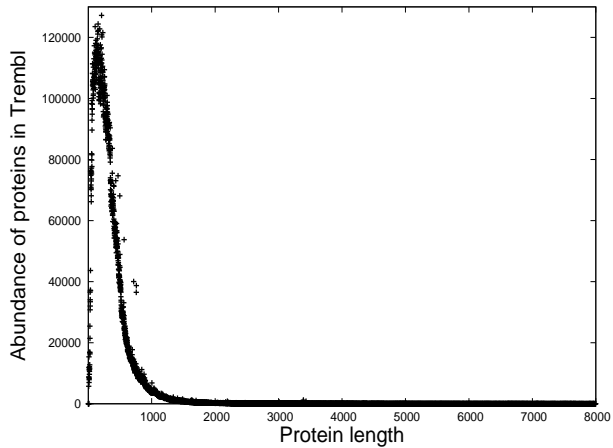}
\caption{A frequency distribution of occurrence rate of proteins versus length of protein in amino acids in TrEMBL v. 18-02.}
\label{fig:trembl_length}
\end{figure}

The CoHSI heterogeneous model has a length distribution with a characteristic form which we exemplify in Fig. \ref{fig:trembl_length}, where we have used the entire version 18-02 of TrEMBL\footnote{ftp://uniprot.org/pub/databases/uniprot/previous\_releases/}. The distribution has a sharply marked unimodal shape which tends to a remarkably accurate power-law for proteins longer than around 200 amino acids.  Precisely the same distribution is found for collections of computer software \cite{HattonWarr2017}.

\begin{figure}[h!]
\centering
\includegraphics[width=8cm,height=6cm]{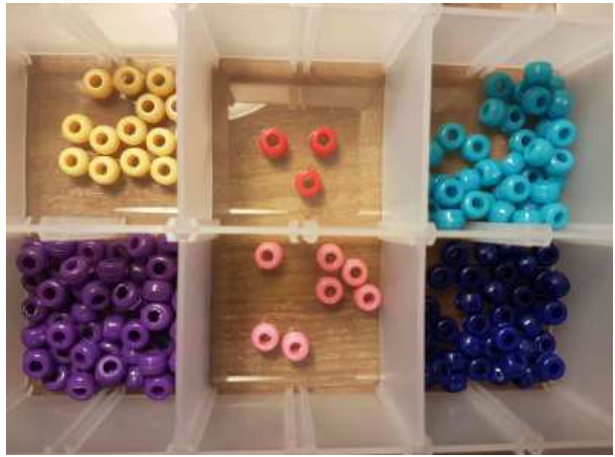}
\caption{Illustrating the CoHSI homogeneous model.  Each box corresponds to a particular number of times a protein is re-used and each of the beads in a box is one of the proteins which is re-used that many times.  Different colors in different boxes simply indicate different levels of re-use.  In this case, the order of beads in the same box is indistinguishable, leading to a different Hartley-Shannon information measure \cite{HattonWarr2017}.}
\label{fig:homogeneous}
\end{figure}

However the same CoHSI theory also demonstrates that heterogeneous behaviour in the sense above can co-exist in the same system with the second \textbf{homogeneous} model for discrete systems, whereby each component contains one particular category, no other component contains the same category and the order of tokens in each component is indistinguishable.  Fig. \ref{fig:homogeneous} illustrates. In this case, CoHSI predicts that homogeneous systems will be overwhelmingly likely to obey a simple power-law \cite{HattonWarr2017}; it is in fact a proof of Zipf's law \cite{Zipf35}.  This too has been validated on multiple qualifying systems to high degrees of confidence \cite{HattonWarr2017}.

Co-existing CoHSI behaviour involving combinations of heterogeneous and homogeneous behaviour turns out to be simply a matter of consistent re-categorization within the same system.  For example, in a text such as a book, by considering the boxes as words and the beads as letters, the distribution of word lengths is a CoHSI heterogeneous system but, in the same text, if we consider boxes as words and the tokens as the frequency with which that word appears, we get a CoHSI homogeneous model which then obeys Zipf's law for the frequency at which words appear \cite{HattonWarr2017}.

\subsubsection*{Choosing a suitable CoHSI model}
\textit{Applying a suitable CoHSI model then amounts to no more than finding a suitable definition or categorization of boxes and beads, and then the CoHSI model leads directly to a predicted length distribution of that categorization.}

We already know that the protein length distribution as a CoHSI heterogeneous model is as shown in Fig. \ref{fig:trembl_length}, but we can also re-categorize our global protein data into notional components such that each component or box contains all those proteins which are re-used $N$ times across the entire known collection of proteins (i.e. the TrEMBL database).  The only requirement is consistency, a \textit{sine qua non} for a sensible experiment.  Clearly no component can be identical to any other since it represents only those proteins with a re-use count of $N$ and the beads are indistinguishable since they only represent those proteins re-used $N$ times without any implication of order.  For example, if $N = 2$, the corresponding component or box contains only those proteins which occur in exactly two separate species across the entire known protein collection, (over 6 million such pairs exist in the case of TrEMBL 18-02, so this box would have over 6 million beads in it).  This simple re-categorization allows us to calculate its Hartley-Shannon information content \cite{HattonWarr2017} as a \textit{homogeneous} model as illustrated by Fig. \ref{fig:homogeneous}.

To make  clear this somewhat abstract process, imagine the following procedure:-

\begin{enumerate}
\item Pick a protein.
\item See how often it is used throughout the TrEMBL release.  For example, suppose this is 10 times, corresponding to the same sequence appearing in 10 different species somewhere in the entire release.
\item Metaphorically paint over the protein to obscure its inner structure and then (metaphorically) paint the number 10 on it and put it in the 10 box with any others. The internal structure of the protein is now irrelevant and is replaced by the property of "10-ness".  This is a consistent re-categorization and each box is unique since if two boxes have the same re-use value, they should simply be merged. In this way it corresponds precisely to the homogeneous model of Fig. \ref{fig:homogeneous}.
\item While there are proteins left, repeat from the first step.
\item When all the proteins have been examined and re-categorized appropriately, count up the contents of each box, arrange them in descending order and by \cite{HattonWarr2017} the prediction is that the data are overwhelmingly likely to obey a power-law.
\end{enumerate}

\textit{In other words, we predict the occurrence rate of protein re-use to follow a power-law, i.e. Zipf's law. Such  \textbf{global}  behavior would ipso facto be independent of phylogeny, natural selection or indeed any mechanism}.  This result would be analogous to the heterogeneous length distribution of proteins shown as Fig. \ref{fig:trembl_length} that can also be explained without recourse to natural selection.  Such outcomes are a direct global consequence of the CoHSI conservation principle acting to guide large discrete systems in which all microstates are equally likely and it is therefore an emergent property.  No mechanism is involved in precisely the same way that no mechanism is involved in the emergence of a lottery winner.  Given the comments about "Darwin is wrong" alluded to earlier, \textit{neither is this inconsistent with natural selection.}  Natural selection simply operates on a system guided on the large-scale by CoHSI in which some global properties have \textit{already} been determined.

So we have a guiding theoretical principle that predicts that the amount of protein re-use is overwhelmingly likely to follow a power-law, irrespective of any mechanism as to how such protein duplication can occur.  In the debate over HGT and VGT, this has the great benefit of being falsifiable.  In order to falsify it however, we need to re-organise the data so that this principle can be tested in reasonable computational time as we have to compare every protein sequence with every other protein sequence in what is currently a 35GB. dataset, even after a winnowing process we now describe.

\subsection*{The protein pipeline}
Guided by CoHSI, of all the parameters available in the full TrEMBL distribution, we are particularly interested in the protein sequence itself, the species and which DL+V it belongs to.  We therefore seek to reduce the entire TrEMBL database to just these factors and we will use the csv (comma-separated variable) file format to represent them.  The rationale behind this is that most programming languages have very simple and efficient ways of handling such files.  Furthermore, providing we order the data appropriately, we can carry out our computations in \textit{one pass} (i.e. without attempting to store intermediate results), by using two venerable but still revolutionary programming technologies: the Unix/Linux \textit{pipeline} \cite{Bourne1983}, and \textit{associative arrays.}  These are both very advantageous for such large files.  We will illustrate use of the pipeline on a case by case basis and we use the programming language \textit{perl} as it handles associative arrays and csv files extremely efficiently.  Associative arrays allow arrays to be indexed by arbitrary strings and we use the obvious choice, the sequence itself.  The fact that some sequences are almost 37,000 amino acids is no barrier to this.

\subsubsection*{Winnowing}
The distributed form of the TrEMBL database contains far more information than we need for this particular task, including publication history, post-translational modifications and many other parameters.  Initial winnowing of the data has been described before in the associated deliverables package for \cite{HattonWarr2015,HattonWarr2016} but in essence using locally written software, we reduce the raw TrEMBL entries during decompression to only those parameters of relevance to this study.  Decompression can be done on the fly in Linux (using built in programs such as \textit{zcat}) so that the original distribution does not have to be explicitly decompressed, (version 18-02 of TrEMBL consumes almost 0.5TB (Terabyte) when decompressed, and the versions are growing rapidly).  This process took approximately an hour on a basic desktop computer with a four core Intel i5-2320 CPU @ 3.00GHz, with 8GB. of memory and a 2TB. disc, and produced a 44 GB. uncompressed csv file \textit{protein\_index.csv} in the following format:-

\begin{verbatim}
 <sequence>,<species name>,<DL+V name>,<length in aa>
\end{verbatim}

This was then sorted into increasing length of sequence as follows and duplicates (i.e. more than one occurrence of the same protein/species) were removed on the fly; this step was necessary as the TrEMBL distribution is automatically annotated without manual review, and the 39 GB. csv file \textit{protein\_index\_sorted.csv}, file was produced, again taking around an hour.

\begin{verbatim}
#1 sort -t"," --key=4 $CSVFILE -n
#2     | uniq -u > protein_index_sorted.csv
\end{verbatim}

\subsubsection*{Data shaping}
After the winnowing stage, the entire TrEMBL database has been reduced by a factor of around 15 and sorted in increasing order on the length of the sequence.  In addition, around 15 million duplicated entries in TrEMBL have been removed by the \textit{uniq} process of line \#2 above leaving just under 92 million protein entries.  Note that both real and virtual (or aggregated) organism codes were used for this as we are only trying to extract individual protein sequences and the species they belong to, (c.f the discussion in \cite{HattonWarr2018c}).  In this form it is now suitable for a highly efficient one-pass treatment (i.e. read only once during processing), and was therefore fed into a locally written small (only 165 lines) data shaping program written in \textit{perl} to reshape the data into the following 34GB. csv format with a file name of \textit{protein\_sharing.csv}

\begin{verbatim}
<length in aa>,<flags>,<aa sequence>,[DL+V name:Species]+,<re-use>
\end{verbatim}

This file contains all the necessary information in a one-pass form for which protein re-use queries can be answered reasonably quickly.

\subsubsection*{Occurrence rate of re-used proteins}
To answer the question as to how often proteins are re-used across species and DL+V, we are now left with a very simple processing pipeline as follows:-

\begin{verbatim}
#1 cat protein_sharing.csv \
#2      | awk '{FS=","}{print $5;}' \
#3      | sort -n \
#4      | uniq -c \
#5      | tail -n +2   \
#6      | ./normalise ccdf   \
#7      > protein_replicas_ccdf.dat
\end{verbatim}

The explanation for each line is

\begin{description}
 \item[\#1] Send the input file down the pipeline.
 \item[\#2] Extract the re-use value losing all other data
 \item[\#3] Sort the stream of re-use values into numerical ascending order
 \item[\#4] Produce pairs consisting of a count along with its re-use value
 \item[\#5] Lose the first line as we only want re-use values which are therefore $> 1$.
 \item[\#6] Turn the data into a ccdf (complementary cumulative distribution function) with a small C program.
 \item[\#7] Place the output into \textit{protein\_replicas\_ccdf.dat} for plotting.
\end{description}

Each row of the \textit{protein\_replicas\_ccdf.dat} file then contains the number of times a protein is re-used followed by the number of proteins involved in that re-use.

\section*{Results - testing  predictions using the full collection of proteins}

\textit{Our straightforward prediction  is that the occurrence rate of protein re-use will follow a power-law, i.e. Zipf's law, and this we test initially using the full TrEMBL release18-02.}

Of the almost 92 million non-duplicate protein entries in TrEMBL 18-02, around 85 million occurred uniquely (i.e. with a re-use count of 1) and these are deliberately omitted as we are interested only in the 7 million or so proteins whose sequences are re-used in more than one species.

\subsection*{The re-use distribution}
\begin{figure}[h!]
\centering
\includegraphics[width=8cm,height=6cm]{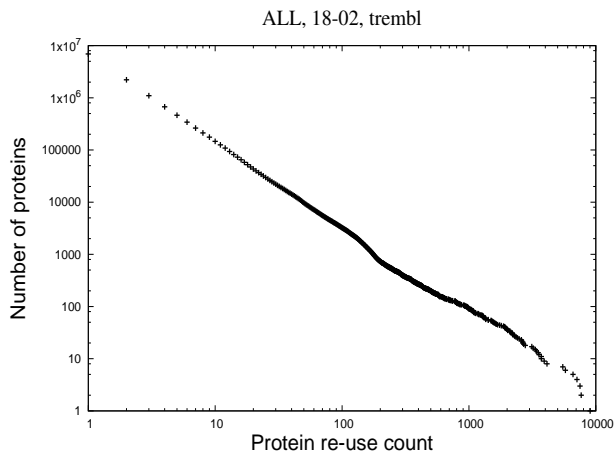}
\caption{A log-log ccdf of the number of re-used proteins and how often they are re-used in all four  domains of life plus viruses (DL+V).}
\label{fig:18-02_trembl_protein_replicas_ccdf}
\end{figure}

This amount of re-use might take the reader by surprise but across the whole of TrEMBL 18-02, the most frequently re-used protein occurred 9,812 times, (in other words exactly the same protein sequence in 9,812 separate species).  In the theory section, we made a prediction using the CoHSI model that the distribution of frequency of protein re-use would be a power-law.  This should appear as a straight line on a log-log ccdf plot \cite{Newman2006}.  Fig. \ref{fig:18-02_trembl_protein_replicas_ccdf} shows that this prediction is extremely accurate, strongly supporting our hypothesis that CoHSI plays a dominant role in protein re-use.

\textit{R lm() reports that the associated p-value matching the power-law linearity in the ccdf of Fig. \ref{fig:18-02_trembl_protein_replicas_ccdf} is $< (2.2) \times 10^{-16}$ over the 4-decade range $1.0-10000.0$, with an adjusted R-squared value of $0.99$.  The slope is $-1.58 \pm 0.20$.}

\subsection*{Accumulating domains in TrEMBL 18-02}
\begin{figure}[h!]
\centering
\includegraphics[width=8cm,height=6cm]{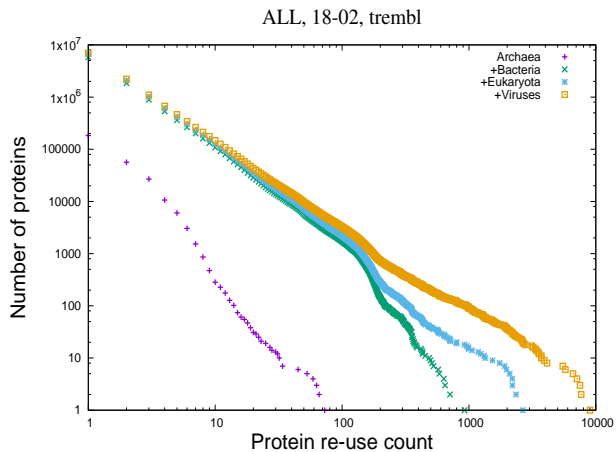}
\caption{An accumulating ccdf of the DL+V in TrEMBL version 18-02.  These show the re-use of archaean proteins (purple); then the re-use of archaean+bacterial proteins (green); then the re-use of archaean+bacterial+eukaryotic proteins (blue) and finally the re-use of archaean+bacterial+eukaryotic+viral proteins (yellow).}
\label{fig:protein_replicas_ccdf_combo}
\end{figure}

We recall that CoHSI predicts an asymptotic Zipf's law \cite{HattonWarr2017}, (i.e. a power-law) for a homogeneous CoHSI model distribution of the number of proteins against the number of times they have been re-used.  Fig. \ref{fig:protein_replicas_ccdf_combo} tells an interesting story.  Here we show as an accumulating system, the re-use of archaean proteins, (relatively a very small contribution); then the re-use of archaean + bacterial proteins; then the re-use of archaean + bacterial + eukaryotic proteins, and finally we include viruses to show the re-use of all proteins in version 18-02 of TrEMBL, including both real and virtual organisms.

\textit{What is apparent is that protein re-use is interdependent with all domains of life and viruses.}  Only if all are included is the predicted asymptotic Zipf's law achieved.  In particular, it appears to depend on the very high re-use rates in viruses to give a full four decade extremely precise power-law.

\begin{figure*}[t!]
    \captionsetup[subfigure]{labelformat=empty}
    \centering
    \begin{subfigure}[t]{0.5\textwidth}
        \centering
        \caption{A}
        \includegraphics[width=6cm]{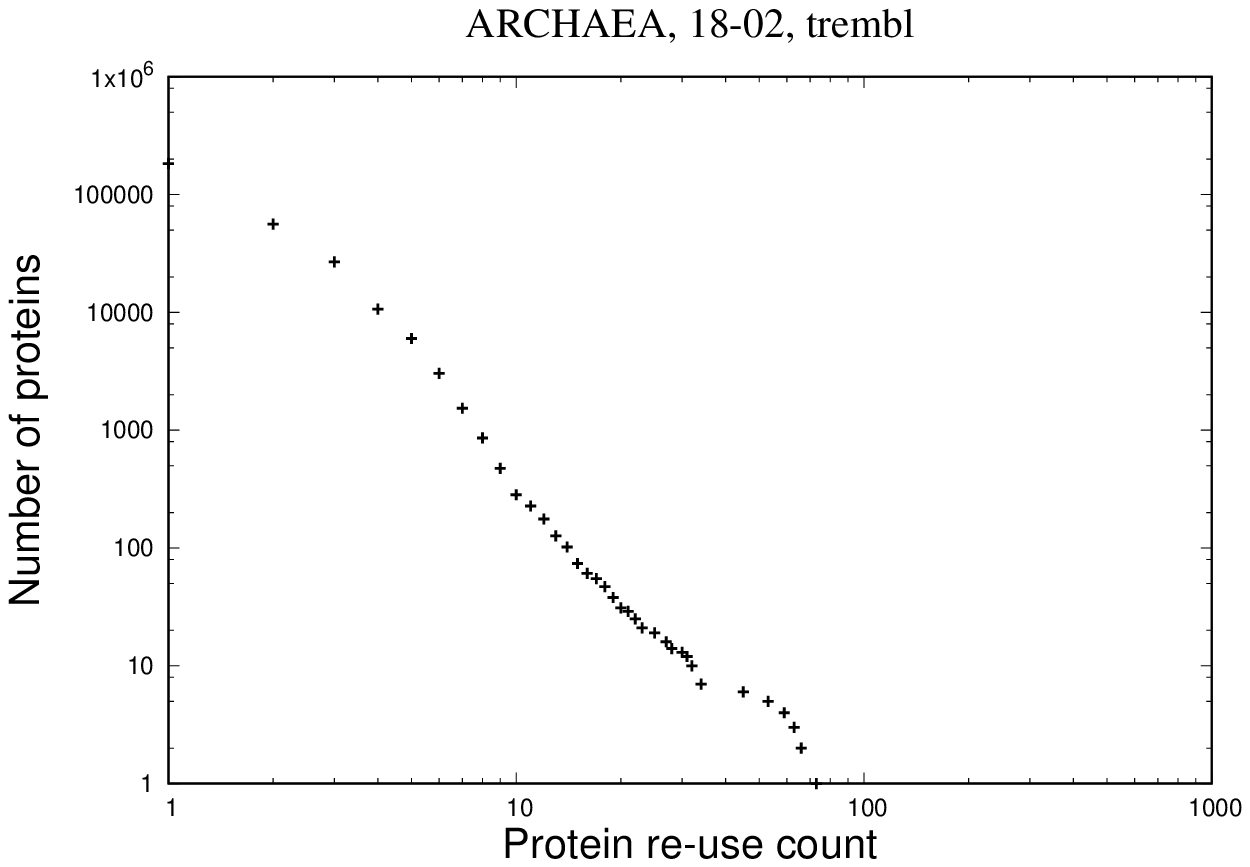}
        \label{fig:18-02_trembl_protein_replicas_ccdf_archaea}
    \end{subfigure}%
    ~
    \begin{subfigure}[t]{0.5\textwidth}
        \centering
        \caption{B}
        \includegraphics[width=6cm]{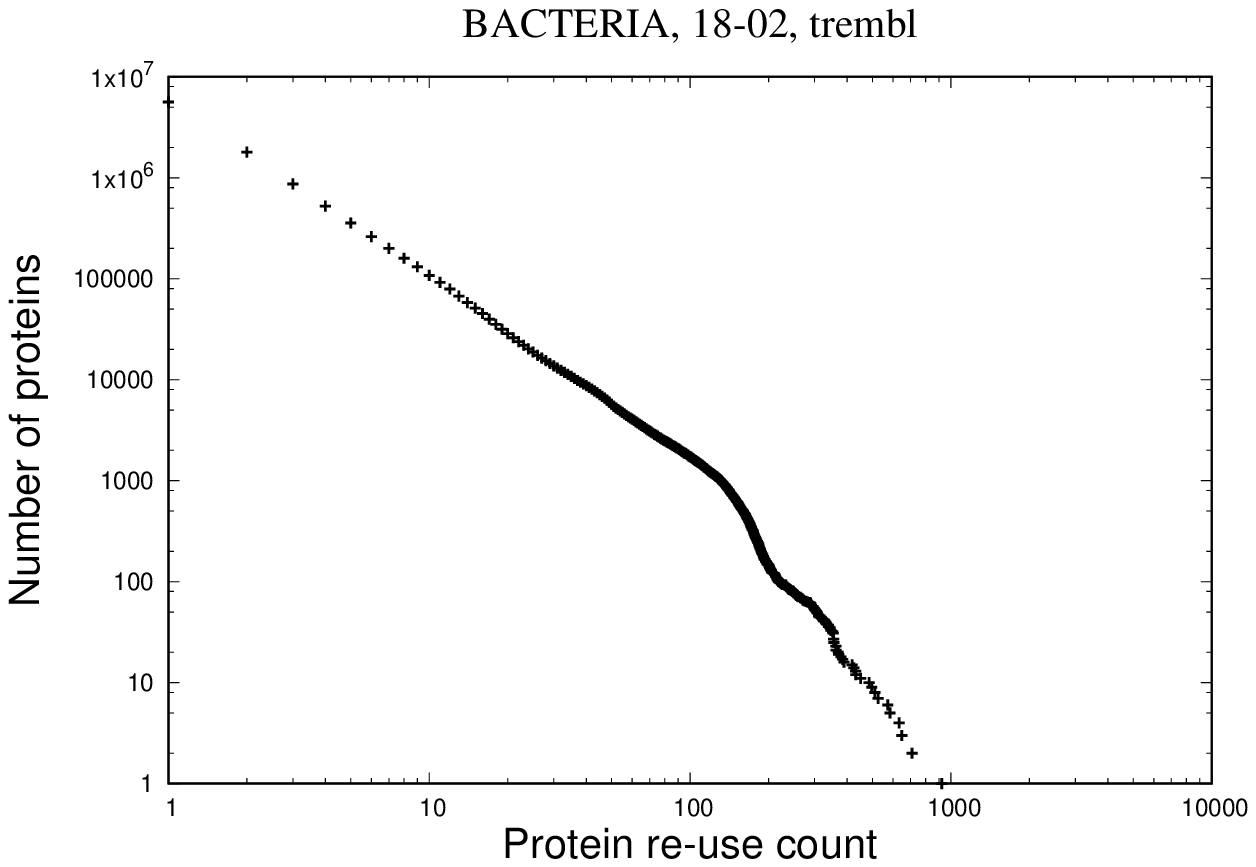}
        \label{fig:18-02_trembl_protein_replicas_ccdf_bacteria}
    \end{subfigure}%

    \begin{subfigure}[t]{0.5\textwidth}
        \centering
        \caption{C}
        \includegraphics[width=6cm]{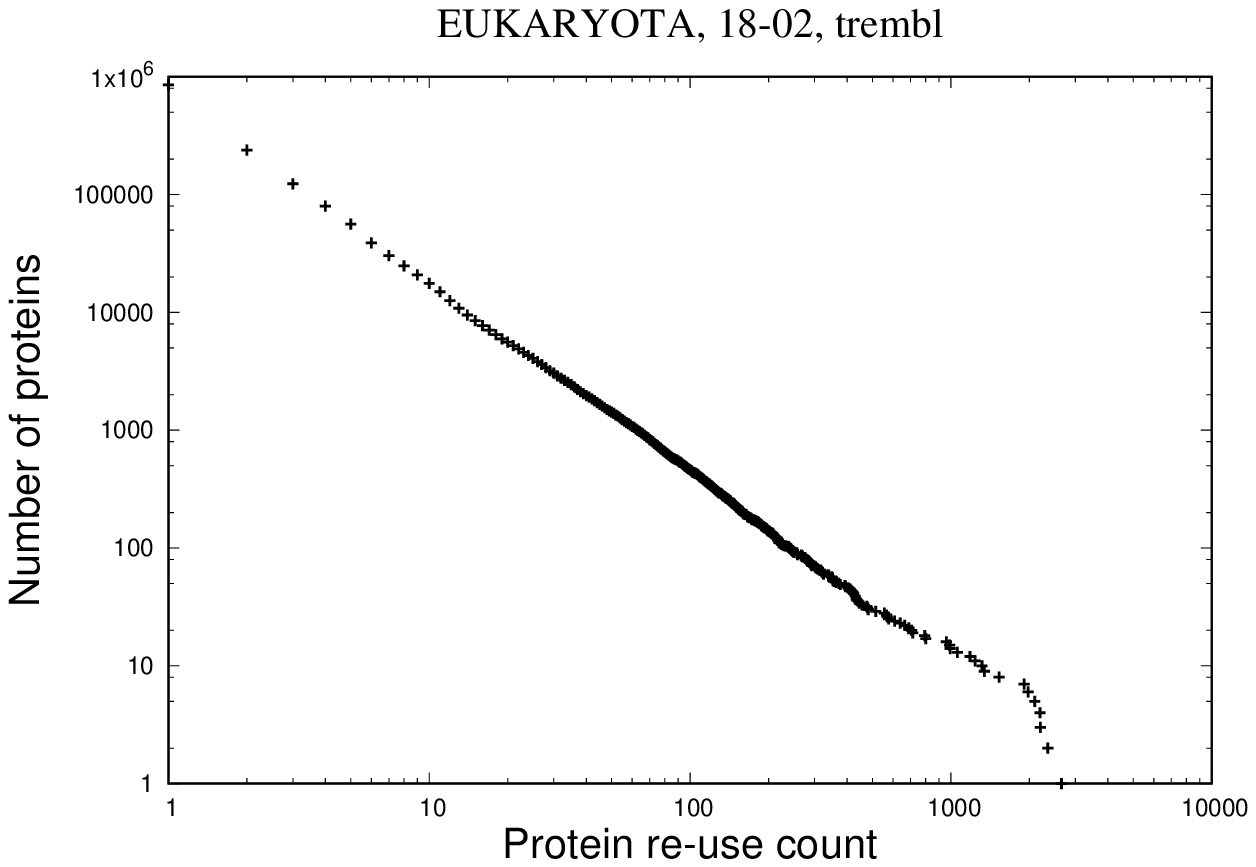}
        \label{fig:18-02_trembl_protein_replicas_ccdf_eukaryota}
    \end{subfigure}%
    ~
    \begin{subfigure}[t]{0.5\textwidth}
        \centering
        \caption{D}
        \includegraphics[width=6cm]{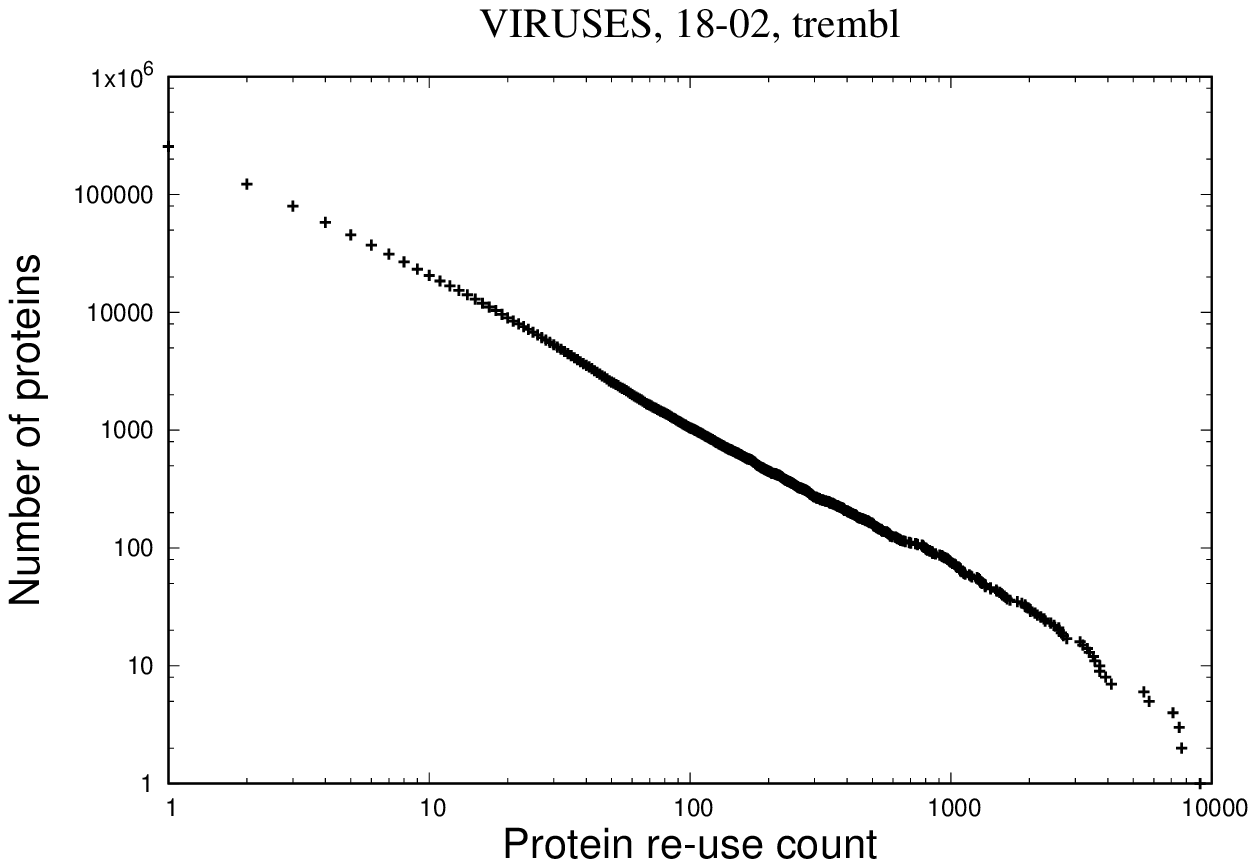}
        \label{fig:18-02_trembl_protein_replicas_ccdf_viruses}
    \end{subfigure}%

    \caption{Each individual domain of life and viruses illustrating the degree of re-use: A) archaea; B) bacteria; C) eukaryota; D) viruses.}
\end{figure*}

In light of this remark, we show in Fig. \ref{fig:18-02_trembl_protein_replicas_ccdf_archaea}-\ref{fig:18-02_trembl_protein_replicas_ccdf_viruses} the contributions of the three domains of life and viruses individually.  The dominant role of the viruses in the highest levels of re-use is obvious, although a small number of proteins are actually involved, (less than 100 or so).  It is also obvious from these figures that archaea, eukaryota and viruses are close to the equilibrium power-law whereas bacteria seem under-represented for re-use greater than around 200.  We will comment on this again shortly.

\subsection*{Results over time for the full TrEMBL datasets }
\begin{figure}[h!]
\centering
\includegraphics[width=8cm,height=6cm]{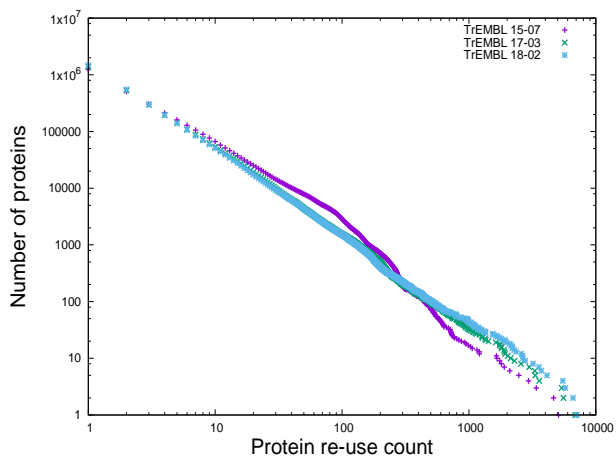}
\caption{A log-log ccdf of the number of re-used proteins and their frequency of re-use  in DL+V taken from 3 releases of the full TrEMBL dataset.}
\label{fig:15-07_18-02_trembl_protein_replicas_ccdf}
\end{figure}

The full unreviewed TrEMBL releases for  2015, 2017 and 2018 were examined and provide an interesting comparison of results as the database not only grew in size over this period  but it was also culled in 2015 and 2016 to remove redundant bacterial sequences. The comparison of TrEMBL releases 15-07, 17-03 and 18-02 is shown in  Fig. \ref{fig:15-07_18-02_trembl_protein_replicas_ccdf}.  This analysis was made more difficult by the availability of only the condensed FASTA format for earlier releases in TrEMBL, but the same property of scale insensitivity is present along with the clear power law relationship. The redundant sequences present in the database before the culling of 2015 and 2016 distorted the protein re-use data and  \textit{the straightening of the power-law lines in post-culling releases (2017 and 2018) is notable.}  

We note in passing that this straightening of Fig. \ref{fig:15-07_18-02_trembl_protein_replicas_ccdf} suggests that the kink in the curve around a re-use of 200 in Fig. \ref{fig:18-02_trembl_protein_replicas_ccdf_bacteria} may be related to curation rather than some intrinsic property of bacteria.

\subsection*{Results over time for a highly curated database }
The SwissProt database includes only those sequences that have been accurately curated manually, and as a result is much smaller (by over two orders of magnitude) than the full TrEMBL database. Thus an examination of multiple releases of both SwissProt and TrEMBL databases is of great interest to see how the predicted power-law patterns might change not only with curation but with the changing size of the release. 

Our initial comparison(Fig. 7) is of three releases of the SwissProt database, the ten year-old 08-09; the 14-02 release; and the 18-02 release. 

\begin{figure}[h!]
\centering
\includegraphics[width=8cm,height=6cm]{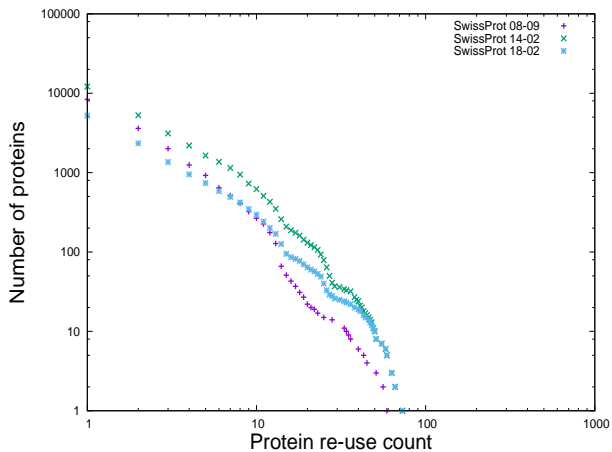}
\caption{A log-log ccdf of the number of re-used proteins in the SwissProt database and how often they are re-used in DL+V.}
\label{fig:08-09_18-02_swissprot_protein_replicas_ccdf}
\end{figure}

Fig. \ref{fig:08-09_18-02_swissprot_protein_replicas_ccdf} shows a log-log ccdf of the re-use in the reviewed releases of SwissProt between 2009 and 2018.  We can note three things: 1) There is  more noise in the SwissProt data than the TrEMBL dataset (cf Fig. 5 and Fig. 6), attributable to the very much smaller size of the SwissProt database. Nevertheless the power law relationship is still clearly visible; 2) the changes as the dataset grows appear to be scale insensitive, (i.e. the plots for all three releases have approximately the same slope, another expected property of CoHSI systems) and 3) the culling of over 50 million redundant bacterial sequences from SwissProt which took place in 2015 and 2016 is easily visible as the 2014 release was bigger than the 2018 release.

\textit{For comparison with the 18-02 TrEMBL release, R lm() reports that the associated p-value matching the power-law tail linearity in the ccdf for the 18-02 SwissProt release shown in Fig. \ref{fig:08-09_18-02_swissprot_protein_replicas_ccdf} is $< (2.2) \times 10^{-16}$ over the 2-decade range $1.0-100.0$, with an adjusted R-squared value of $0.95$.  The slope is $-1.88 \pm 0.40$.}

\subsection*{Local and extended protein re-use}
At this stage, we now want to access re-use information in detail. In line with our approach above, if we wish to search for different patterns, it is necessary to re-shape the data into a different format to make such comparisons computationally efficient.  For re-use analysis, we choose the csv format file described in the data shaping section which has the following fields:-

\begin{verbatim}
<aa>,<flags>,<sequence>,<DL+V name:Species>,<re-use>
\end{verbatim}

The flags field consists of up to 4 ``*'' indicating whether the particular protein occurs across DL+V and in how many of these.  4 ``*'' therefore indicates a protein which occurs in species in archaea, bacteria, eukaryota AND viruses.

The fourth field consists of multiple occurrences of ``DL+V name:Species'' separated by the ``/'' character indicating where the specific protein is re-used.

Some examples of proteins shared across different DL+V are shown in Table \ref{tab:reuse}.  Note that for clarity we show only the "real" protein entries and not the "virtual" protein entries which begin with the code "9", which appear in the TrEMBL database, but not in the SwissProt database.

\begin{table}[!ht]
\centering
\begin{tabular}{p{1cm}p{1cm}p{5cm}p{4cm}p{1cm}}
\hline
aa & flags & sequence & re-used in & count \\
\hline
15 & **& \makecell{ANDNFAAEGDVAVAA} & \makecell{EUKARYOTA:BRUMA \\ BACTERIA:WOLTR} & 2 \\
\hline
23 & **& \makecell{RIVATAASRSTCKFA \\ SSSTTHIK } & \makecell{VIRUSES:SCVLA \\ EUKARYOTA:YEASX} & 2 \\
\hline
38 & **& \makecell{MTALIRHWEKWSGWY \\ LFLTAVSAWLYLLAV \\ IFREGWIR} & \makecell{VIRUSES:BPE15 \\ BACTERIA:CITFR} & 2 \\
\hline
48 & **& \makecell{MMLMYQCLRCGSIFD \\ KRSEVIEHLLSVHGQ \\ MNKVTLEYFYIYFKV \\ RRP} & \makecell{VIRUSES:SSV2 \\ ARCHAEA:SULSF} & 2 \\
\hline
55 & **& \makecell{MKMPEKHDLLAAILA \\ AKEQGIGAILAFAMA \\ YLRGRYNGGAFTKTV \\ IDATMCAIIA} & \makecell{BACTERIA:BACTU \\ ARCHAEA:METSM} & 2 \\
\hline
\end{tabular}
\caption{Some examples of re-used proteins in Trembl 18-02.}
\label{tab:reuse}
\end{table}

We summarise the overall protein re-use in Table \ref{tab:resusecount}
\begin{table}[!ht]
\centering
\begin{tabular}{p{6cm}r}
\hline
Number of DL+V present & Re-use count \\
\hline
1 & 6,906,551 \\
2 & 21,676 \\
3 & 22 \\
4 & 5 \\
\hline
\end{tabular}
\caption{Some examples of re-use in Trembl 18-02.}
\label{tab:resusecount}
\end{table}

The number of protein entries in each DL+V is given by Table \ref{tab:lgt0}.
\begin{table}[!ht]
\centering
\begin{tabular}{p{6cm}r}
\hline
DL+V & Protein count \\
\hline
ARCHAEA & 2,016,539 \\
BACTERIA & 57,922,123 \\
EUKARYOTA & 22,951,472 \\
VIRUSES & 1,805,747 \\
\hline
\end{tabular}
\caption{Protein appearance by DL+V in Trembl 18-02.}
\label{tab:lgt0}
\end{table}

Breaking Table \ref{tab:resusecount} down into individual protein re-use is shown in the following tables.  Table \ref{tab:lgt1} shows how many re-used proteins appeared for each DL+V. 

\begin{table}[!ht]
\centering
\begin{tabular}{p{6cm}r}
\hline
DL+V & Protein re-use count \\
\hline
ARCHAEA and ARCHAEA & 181,427 \\
BACTERIA and BACTERIA & 5,628,681 \\
EUKARYOTA and EUKARYOTA & 853,386 \\
VIRUSES and VIRUSES & 243,057 \\
\hline
\end{tabular}
\caption{Protein re-use by DL+V in Trembl 18-02.}
\label{tab:lgt1}
\end{table}

\begin{table}[!ht]
\centering
\begin{tabular}{p{6cm}r}
\hline
DL+V pairs & Protein re-use count \\
\hline
ARCHAEA and BACTERIA & 3,530 \\
ARCHAEA and EUKARYOTA & 6 \\
ARCHAEA and VIRUSES & 201 \\
BACTERIA and EUKARYOTA & 2,844 \\
BACTERIA and VIRUSES & 14,824 \\
EUKARYOTA and VIRUSES & 271 \\
\hline
\end{tabular}
\caption{Protein re-use in pairs of DL+V in Trembl 18-02.}
\label{tab:lgt2}
\end{table}

\begin{table}[!ht]
\centering
\begin{tabular}{p{6cm}r}
\hline
DL+V triplets & Protein re-use count \\
\hline
ARCHAEA and BACTERIA and EUKARYOTA & 1 \\
ARCHAEA and BACTERIA and VIRUSES & 1 \\
ARCHAEA and EUKARYOTA and VIRUSES & 0 \\
BACTERIA and EUKARYOTA and VIRUSES & 20 \\
\hline
\end{tabular}
\caption{Protein re-use in triplets of DL+V in Trembl 18-02.}
\label{tab:lgt3}
\end{table}

\begin{table}[h!]
\centering
\begin{tabular}{p{6cm}r}
\hline
DL+V quadruplets & Protein re-use count \\
\hline
ARCHAEA and BACTERIA and EUKARYOTA and VIRUSES & 5 \\
\hline
\end{tabular}
\caption{Protein re-use in quadruplets of DL+V in Trembl 18-02.}
\label{tab:lgt4}
\end{table}

At this point, the value of data shaping and pipelining becomes fully apparent with Tables \ref{tab:reuse}, \ref{tab:resusecount}, \ref{tab:lgt1}-\ref{tab:lgt4} taking only a few minutes to produce.

\subsection*{Sensitivity analysis}
Small variations in sequence between proteins have two potential origins. They  arise naturally by genetic mechanisms in the course of evolution (e.g. as alleles, paralogs and orthologs) but can also result from experimental error. An examination of the data cannot distinguish \textit{prima facie} from which of these causes any identified variations have arisen, and thus the enumeration of sequence variants will place an upper bound on the potential impact on our results of experimental error. 

The structure of the dataset made it feasible to check all approximately 92 million protein sequences for lexical closeness to their neighbours when sorted in increasing length in amino acids, of the following kinds:-

\begin{description}
 \item[Transposition] A difference in sequence caused by a single transposition of adjacent aa.
 \item[One letter insertions, deletions or substitutions] A difference of one aa at any position, an extra aa, or a missing aa.
\end{description}

To carry out this analysis efficiently requires sorting the protein data into increasing sequence length so that the comparisons can be done in one pass.  Furthermore, we must consider \textit{all} the proteins in the 18-02 distribution and not just the re-used ones as one of these potential sequence errors could cause a non re-used protein to appear as a re-used protein, and \textit{vice versa}.  The results (which took less than an hour using a fast lexical string checker written in C and included in the reproducibility deliverables), are shown as Table \ref{tab:lexical}.

\begin{table}[h!]
\centering
\begin{tabular}{p{6cm}rr}
\hline
Type of lexical similarity & count & Percentage of total \\
\hline
Identical apart from transpositions & 5,226 & 0.006\% \\
Identical apart from one position difference & 4,166,632 & 4.5\% \\
\hline
\end{tabular}
\caption{Potential intrusion of experimental error due to close lexical similarity in protein sequence in an analysis of 91,624,134 proteins with unique entries in the TrEMBL database release 18-02.}
\label{tab:lexical}
\end{table}

As can be seen, the maximum potential impact of experimental error in so far as it might cause transpositions or single amino acid insertions, deletions or substitutions is less than 5\% of the dataset and we therefore consider it unlikely to have any significant effects on these results presented here and their conclusions.

\section*{Conclusions}
Based on a conservation principle (CoHSI) derived from information theory embedded as a constraint within a Statistical Mechanics methodology, we have proposed a novel explanation of the frequency of gene transfer (inferred from encoded proteins), that is independent of mechanism (i.e. HGT or VGT).  Our explanation exploits the properties of the token- and mechanism-agnostic CoHSI methodology and unites HGT and VGT under a single umbrella term, \textit{protein re-use.}  Using this methodology  we predicted that identical protein copies will be re-used (i.e.distributed across the phylogenetic tree) as a power-law in the frequency at which they are re-used.  We have then demonstrated by extensive analysis of the TrEMBL database that

\begin{enumerate}
\item The predicted power-law in frequency of protein re-use is confirmed with extremely high statistical significance, (linearity on a log-log ccdf, adj. $R^{2} = 0.99, p < 2.2 \times 10^{-16}, slope = -1.58 \pm 0.2$ over 4 decades).  As in \cite{HattonWarr2018c}, this slope appears to be robust with respect to database size implying that as the database grows, the maximum amount of re-use will grow in an entirely systematic manner.
\item This property owes nothing to natural selection or indeed any mechanism.  None is required - it is simply an emergent property of a CoHSI homogeneous system, Fig. \ref{fig:18-02_trembl_protein_replicas_ccdf}, just as the length distribution of proteins in amino acids is a natural property of a CoHSI heterogeneous system, Fig. \ref{fig:trembl_length}.
\item Viruses play an essential role in the establishment of this emphatic power-law behaviour, Fig. \ref{fig:protein_replicas_ccdf_combo}, by contributing a disproportionate number of highly re-used proteins which carry the power-law into 4 decades.  Indeed, from the point of view of information theory, there is no distinction between any of the three domains of life and viruses in this regard.
\item Proteins are re-used as often as 9,812 times and re-used proteins can be as long as 14,596 amino acids.
\item Our result is robust with respect to potential experimental errors leading to single amino acid polymorphisms in the form of changes, insertions or deletions, or to transpositions between adjacent amino acids.
\item In terms of data-processing, we have demonstrated the importance of simple data re-shaping in order to query the immense and growing protein sequence resources in reasonable time.  Unusual queries are not generally well-handled by conventional databases and yet unusual queries might ask very revealing questions.
\end{enumerate}

We conclude by returning to the profound impact first made by the discovery of HGT which prompted the headline "Darwin is wrong" referenced in the introduction.  We have demonstrated repeatedly in a number of previous papers that CoHSI is a deeper principle than natural selection in that it shapes the landscape on which natural selection operates.  In other words, it is entirely complementary to natural selection, relieving natural selection from the burden of utilizing an unedifying thicket of local explanations which may be difficult or indeed impossible to falsify in attempting to explain global properties of a system.

Instead, we have demonstrated by testing a falsifiable prediction that HGT and  VGT can be unified under the umbrella of \textit{protein re-use}, which is a global property of a discrete system shaped by a conservation principle and the CoHSI distribution.  In light of this finding, the assertion "Darwin was wrong" is beside the point. With respect to protein re-use,  this phenomenon operates at a deeper level than natural selection.

\section*{Acknowledgements}
We would like thank Gillian Libretto for insights into some of these issues and for bringing the \textit{New Scientist} article to our attention.

\begin{itemize}
 \item \textbf{Correspondence:} Correspondence and requests for materials
should be addressed to Les Hatton ~(email: lesh@oakcomp.co.uk).
\end{itemize}

\section*{Author's contributions}
LH performed the analyses, LH and GW developed the arguments, discussed the results and contributed to the text of the manuscript.

\section*{Competing interests}
The authors declare no competing financial interests.

\section*{Funding}
This work was unfunded.

\newpage


%
%

\bibliographystyle{alpha}
\bibliography{bibliography}

\end{document}